\pdfoutput=1
\def\year{2020}\relax
\documentclass[letterpaper]{article} % DO NOT CHANGE THIS

\usepackage{aaai20}  % DO NOT CHANGE THIS
\usepackage{times}  % DO NOT CHANGE THIS
\usepackage{helvet} % DO NOT CHANGE THIS
\usepackage{courier}  % DO NOT CHANGE THIS
\usepackage[hyphens]{url}  % DO NOT CHANGE THIS
\usepackage{graphicx} % DO NOT CHANGE THIS
\usepackage{amsfonts}
\usepackage{algorithm}
\usepackage{algorithmic}
\usepackage{amsmath}
\usepackage{amssymb}
\usepackage{amsfonts}
\usepackage{bm}
\usepackage{color}
\usepackage{comment}
\usepackage{graphicx}
\usepackage{multirow}
\usepackage{subfigure}

\newcommand{\Mat}[1]{\mathbf{#1}}

\usepackage{cleveref}

\usepackage{graphicx}  % DO NOT CHANGE THIS

\title{Revisiting Graph based Collaborative Filtering: A Linear Residual Graph Convolutional Network Approach}

\author {
Lei Chen\textsuperscript{\rm 1,2}, 
Le Wu\textsuperscript{\rm 1,2,}\thanks{Corresponding Author}, 
Richang Hong\textsuperscript{\rm 1,2}, 
Kun Zhang\textsuperscript{\rm 3},
Meng Wang \textsuperscript{\rm 1,2} \\
\textsuperscript{\rm 1}Key Laboratory of Knowledge Engineering with Big Data, Hefei University of Technology \\
\textsuperscript{\rm 2}School of Computer Science and Information Engineering, HeFei University of Technology \\
\textsuperscript{\rm 3}School of Computer Science and Technology, University of Science and Technology of China\\ 
\{chenlei.hfut,lewu.ustc, hongrc.hfut\}@gmail.com, 
zhkun@mail.ustc.edu.cn,
eric.mengwang@gmail.com 
}

\begin{document}

\maketitle

\begin{abstract}

Graph Convolutional Networks~(GCNs) are state-of-the-art graph based representation learning models by iteratively stacking multiple layers of convolution aggregation operations and non-linear activation operations. Recently, in Collaborative Filtering~(CF) based Recommender Systems~(RS), by treating the user-item interaction behavior as a bipartite graph, some researchers model higher-layer collaborative signals with GCNs. These GCN based recommender models show superior performance compared to traditional works. However, these models suffer from training difficulty with non-linear activations for large user-item graphs. Besides, most GCN based models could not model deeper layers due to the over smoothing effect with  the graph convolution operation. In this paper, we revisit GCN based CF models from two aspects.  First, we empirically show that removing non-linearities would enhance recommendation performance, which is consistent with the theories in simple graph convolutional networks. Second, we propose a residual network structure that is specifically designed for CF with user-item interaction modeling, which alleviates the over smoothing problem in graph convolution aggregation operation with sparse user-item interaction data. The proposed model is a linear model and it is easy to train, scale to large datasets, and yield better efficiency and effectiveness on two real datasets. We publish the source code at https://github.com/newlei/LR-GCCF.

\end{abstract}

\section{Introduction}

Recent years have witnessed the boom of GCNs, which are efficient variants of CNNs for dealing with graph based data~\cite{ICLR2017semi,NIPS2017inductive}. The key idea of GCNs is to stack multiple layers that iteratively perform the following two steps at each layer: node embedding with convolutional neighborhood aggregation; followed by a non-linear transformation of node embeddings parameterized by a neural network. Therefore, the higher-order similarity of a node could be effectively captured~\cite{ICLR2019powerful,li2018deeper}. These models show competing performance for tasks such as unsupervised node~(graph) representation learning~\cite{li2016unsupervised}, semi-supervised node~(graph) learning~\cite{ICLR2017semi,camps2007semi}, and so on.

As many real-world data show the graph structure, GCNs have been widely applied to applications such as social network analysis~\cite{xu2019relation}, transportation network~\cite{zhao2019t}, and recommender systems~\cite{kdd2018GCMC}. In this paper, we focus on applying GCNs to CF based recommender systems. CF provides personalized item suggestions to users by learning user and item embeddings from their historical behavior data~\cite{UAI2009bpr,WWW2017neural}. In fact, by treating the user-item historical behavior as a bipartite graph with edges between users and items, CF can be naturally transformed into the edge prediction problem in the graph. This graph representation of user-item behavior advances previous user-item 
6
 interaction matrix with more higher-order user and item correlations, and provides the possibility to alleviate the data sparsity issue in CF with graph structure modeling~\cite{sigir2019ngcf,kdd2018GCMC,kdd2018Pinsage}. Some earlier works applied personalized random walks~\cite{SIGIR2008eigenrank} or relied on graph regularization models with auxiliary graph data~(e.g., social network) for recommendation~\cite{SDM2010collaborative,huang2004graph}. These models suffered from a huge time complexity with personalized random walk, and most of these models relied on carefully designing the random walk process. Recently, plenty of researchers pay more attention  to apply  
GCNs for recommendation~\cite{kdir18,sigir2019ngcf,kdd2018GCMC,kdd2018Pinsage}. For example, PinSage designed sampling techniques for graph convolution aggregation to alleviate the computational burden in the recommendation process~\cite{kdd2018Pinsage}. By feeding the user and item free embeddings as input, NGCF was specially designed for GCN based CF~\cite{sigir2019ngcf}. NGCF iteratively propagates user and item embeddings in the graph to distill the collaborative signals with graph convolutions.
These GCN based recommender models show better performance compared to traditional models.

Despite the relative success of GCN based recommendation, we argue that two important problems in GCN based CF still remain unsolved. On one hand, for user and item embeddings, GCNs follow the two steps of neighborhood aggregation with graph convolutional operations and non-linear transformations. While graph convolutional operations are effective for aggregating the neighborhood information and modeling higher order graph structure, is the additional complexity introduced by the non-linear feature transformation in GCNs necessary? On the other hand, most of the current GCN based models could only stack very few layers~(e.g., 2 layers).  In fact, the graph convolution operation is a special kind of Graph Laplacian smoothing~\cite{li2018deeper,klicpera2018predict}. With $K$-th layer of GCNs, the Laplacian smoothing is performed to incorporate the up to $K$-th neighbors. Therefore, the over-smoothing effect exists with deep layers, as the higher layer neighbors tend to be indistinguishable for each node. With limited user-item interaction records in the recommendation~\cite{wu2017modeling,wu2016relevance}, this problem would become more severe since the training records are very sparse. Intuitively, with the increasing of the stacking layers, the smoothing effect could alleviate the data sparsity of CF at first, but the over smoothing effect introduced by more layers would neglect each user's uniqueness and degrade the recommendation performance. How to better model the graph structure while avoiding the over smoothing effect in this process remains pretty much open.

To tackle the above two issues, we revisit the graph based CF models with a linear residual graph convolutional approach. Our main contributions lie in two aspects: on one hand, we empirically analyze the uniqueness of CF from most graph based tasks, and show that removing the nonlinearity would enhance the recommendation performance with less complexity, which is consistent with the recent theories in simplifying GCNs~\cite{ICML2019simplifying}. Furthermore, to alleviate the over smoothing problem in the iterative process, we propose to learn the residual user-item preference at each layer. Thus, the user uniqueness is preserved at the lower layers, while the higher layers of the GCNs could focus on learning users' residual preferences that could not be captured from each user's limited historical records. Please note that this idea is inspired the ResNet architecture in CNNs~\cite{CVPR2016resudial,wu2019wider}, and our work focuses on how to extend the formulation of the residual part in CF with the interaction prediction between users and items under GCNs. We then show that with linear residual learning, our proposed model degenerates to a linear model that effectively leverages the user-item graph structure for recommendation. In summary, in contrast to current GCN based recommendation models, our proposed model is easier to train, scales to large datasets. Finally, we perform extensive experiments on two large real-world CF datasets, and the results clearly show the effectiveness and efficiency of our proposed model.

\section{Preliminaries and Related Work}
Considering a graph $\mathcal{G}\!=\!<\!\mathcal{V}, \!\mathbf{A}>$, with $\mathcal{V}$ is the set of nodes and $\mathbf{A}$ is the adjacency matrix, in which $a_{ij}$ denotes the edge between node $i$ and node $j$. If there is a directed edge from node $i$ to node $j$, then $a_{ij}\!=\!1$, otherwise it is 0. For ease of notation, we use $S_i\!=\![j| a_{ij}\!=\!1]$ to denote the neighbor set of node $i$, i.e., the node set that $i$ connects to. We use $\Mat{S}\!=\!\Mat{\tilde{D}}^{-0.5}\Mat{\tilde{A}}\Mat{\tilde{D}}^{-0.5}$ to denote the normalized adjacency matrix with added self loops, with $\Mat{\tilde{A}}=\Mat{A}+\Mat{I}$ is the adjacency matrix of the graph with added self-connections, and~$\mathbf{I}$ is the identity matrix.~$\Mat{\tilde{D}}$ is the degree matrix of~$\Mat{\tilde{A}}$.

\subsection{Graph Convolutional Networks}
For each node $i\!\in\!\mathcal{V}$, we use $\mathbf{h}^0_i$ to denote the node initial embedding, which is usually the feature vector $\mathbf{x}_i$ of node $i$~(i.e, $\mathbf{h}^0=\mathbf{x}_i$).  In a graph $\mathcal{G}$, the key idea of GCNs is to stack $K$ steps in a recursive message passing or feature propagation manner to learn node embeddings~\cite{kdd2018GCMC,NIPS2017inductive,ICML2017neural}. Specifically, for each node $i$ at the step, it is computed recursively with following two steps: feature propagation and non-linear feature transformation.

\textbf{Feature propagation.}  
For each node $i$, the feature aggregation step aggregates the embeddings from graph neighbors $S_i$ and its own embedding $\mathbf{h}^k_i$ at previous layer $k$.
 
Earlier works focus on how to model the aggregation functions~\cite{NIPS2017inductive,ICML2017neural,ICLR2018graphATT,ICLR2017semi}. As the focus of this paper is not to design more sophisticated feature aggregation function, we follow the widely used feature aggregation function proposed in Kipf et al.~\cite{ICLR2017semi}, which is empirically effective and has been adopted by many GCN variants~\cite{ICLR2017semi,kdd2018GCMC,ICML2019simplifying}:

\begin{small} 
\begin{equation}\label{eq:gcn_fpw_matrix}
\bar{\Mat{H}}^{(k+1)}=\Mat{\tilde{D}}^{-0.5}\Mat{\tilde{A}}\Mat{\tilde{D}}^{-0.5}\Mat{H}^k.
\end{equation} 
\end{small}

In fact, given the features $\Mat{H}^k$ at $k$-th layer, feature propagation output $\bar{\Mat{H}}^{(k+1)}$
layer can be regarded as the Laplacian smoothing on the features at the previous layer~\cite{li2018deeper,zhu2003semi}.

\textbf{Nonlinear transformation.} The nonlinear transformation layer is a standard Multilayer Perceptron~(MLP). By feeding the output of the feature propagation step, the nonlinear transformation produces the $(k+1)$-th layer embedding of each node as:

\begin{small}
\vspace{-0.3cm}
\begin{equation} \label{eq：gcn_ntrans}
\mathbf{H}^{(k+1)}=\sigma(\bar{\Mat{H}}^{(k+1)}\mathbf{W}^k),
\end{equation}
\vspace{-0.4cm}
\end{small}

\noindent where $\sigma(x)$ is a non-linear activation function.

After iteratively performing the two steps in each layer with a defined depth $K$, the final embedding of each node at depth $K$ is~{\footnotesize $\mathbf{h}^K_i$}. For most GCN based applications, there is a prediction function $f(.)$ as:

\begin{small}
\vspace{-0.3cm}
\begin{equation}
\hat{y}= f({\mathbf{h}^K_i|i\in V}).
\end{equation}
\vspace{-0.3cm}
\end{small}

As GCNs derive inspiration primarily from the CNNs in the deep learning community, it inherits considerable nonlinearity and complexity from the nonlinear transformations as shown in Eq.\eqref{eq：gcn_ntrans}. Researchers exploit the possibility of simplifying
GCNs. Recently, a Simple Graph Convolution~(SGC) is proposed~\cite{ICML2019simplifying}, which removes the nonlinear transformation in Eq.\eqref{eq：gcn_ntrans} as:

\begin{small}
\vspace{-0.cm}
\begin{equation} \label{eq：sgcn_1}
\mathbf{H}^{K}=\mathbf{\mathbf{S}\mathbf{S}...\mathbf{S}\mathbf{H^0}\Mat{W}^0\Mat{W}^1...\Mat{W}^K},
\end{equation} 
\end{small}

\noindent where we can rewrite~{\small$\Mat{W}^0\Mat{W}^1...\Mat{W}^K$} as a single matrix~$\Mat{W}$, and the above linear matrix multiplication turns to:

\begin{small}
\vspace{-0.2cm}
\begin{equation} \label{eq：sgcn_1}
\mathbf{H}^{K}=\mathbf{\mathbf{S}\mathbf{S}...\mathbf{S}\mathbf{H^0}\Mat{W}}.
\end{equation} 
\end{small}

With the formulation of SGC, GCNs reduce to the iterative simple feature propagations with very few parameters. Therefore, it is easy to tune and scales to large datasets. As verified by researchers, SGCN corresponds to a fixed low pass filter on graph spectral domain. Besides, the empirical evaluations show that SGCN does not negatively impact accuracy in many graph based tasks with huge time improvement~\cite{ICML2019simplifying}.

\subsection{Graph Convolutional based Recommendation}
In a recommender system, there are two sets of entities: a userset~{\small$\mathcal{U}$} with {\small$M$} users~({\small{$|\mathcal{U}|\!=\!M$}})~and an itemset {\small$\mathcal{V}$}~({\small{$|\mathcal{V}|\!=\!N$}}). As implicit feedback is the most common form in many recommender systems, we focus on implicit feedback based CF in this paper~\cite{UAI2009bpr}, and it is easy to extend the proposed model for rating prediction in CF.
Users show ratings to the items with a rating matrix~{\small$\Mat{R}\in \mathbb{R}^{M\times N}$}, with $r_{ai}\!=\!1$  denotes user $a$ likes item $i$, otherwise it equals 0. With the rating matrix, accurately learning user embedding matrix and item embedding matrix is a key to the success of recommendation performance. Earlier works focus on shallow matrix factorization based models~\cite{koren2009matrix,UAI2009bpr}. Deep learning based models, e.g., NeuMF~\cite{WWW2017neural}, and Wide\&Deep~\cite{recsys2016wide} modeled the interaction between users and items with a deep neural network structure.

With the huge success of GCNs, researchers attempted to formulate recommendation as a user-item bipartite graph, and adapted GCNs for recommendation~\cite{sigir2019ngcf,NIPSi2017geometric,kdd2018Pinsage}. Earlier works on GCN based models relied on the spectral theories of graphs, and are computationally costly when applying in real-world recommendations~\cite{NIPSi2017geometric,zheng2018spectral}. Some of recent works on GCN based recommendation models focused on the spatial domain~\cite{kdir18,sigir2019ngcf,kdd2018GCMC,kdd2018Pinsage}. PinSage was designed for similar item recommendation under the content based model, with the item features $\mathbf{x}_v$ and the item-item correlation graph as the inputs~\cite{kdd2018Pinsage} . GC-MC~\cite{kdd2018GCMC} and NGCF~\cite{sigir2019ngcf} are specifically designed under the CF setting. Given ratings of users to items, the user-item bipartite graph is denoted as
{\small$\mathcal{G}\!=\!<\mathcal{U}\cup \mathcal{V}, \mathbf{A}>$}, with {\small$\mathbf{A}$} is constructed from the rating matrix {\small$\Mat{R}$} as:

\begin{small}
\vspace{-0.2cm}
\begin{equation} \label{eq:cfmat2graph}
\mathbf{A}=\left [
    \begin{array}{c c}\Mat{R} \quad& \Mat{0}^{N\times M} \\
    \Mat{0}^{M\times N}\quad&\Mat{R^T} 	
    \end{array}
    \right].
\end{equation} 
\end{small}

Let {\small$\Mat{E}\in\mathbb{R}^{(M+N)\times D}$} denote the free embedding matrix of users and items. By feeding the free embedding matrix  {\small$\Mat{E}$} into GCNs with bipartite graph $\mathcal{G}$, i.e., $\small{\forall i\in \mathcal{U}\cup \mathcal{V}}, \mathbf{h}^0_i=\mathbf{e}_i$. Then, GCNs iteratively perform with embedding propagation step in Eq.\eqref{eq:gcn_fpw_matrix} and nonlinear transformation with Eq.\eqref{eq：gcn_ntrans} and each user's~(item's) embeddings can be updated in the iterative process. Therefore, the final embedding {\footnotesize $\Mat{H}^K$} explicitly injects the up to $K$-th order collective connections between users and items.
All the parameters~(including the initial free embedding matrix~{\small$\Mat{E}$}, the transformation parameters~({\small $[\Mat{W}^k]_{k=0}^K$})) can be learned in an end-to-end manner. GC-MC could be seen a special case of NGCF with $K=1$, i.e., only the first order connectivity of the user-item bipartite graph is modeled~\cite{kdd2018GCMC}.

\begin{small}
\vspace{-0.2cm}
\begin{figure*} [htb]
	\begin{center}
		\includegraphics[width=2.0\columnwidth]{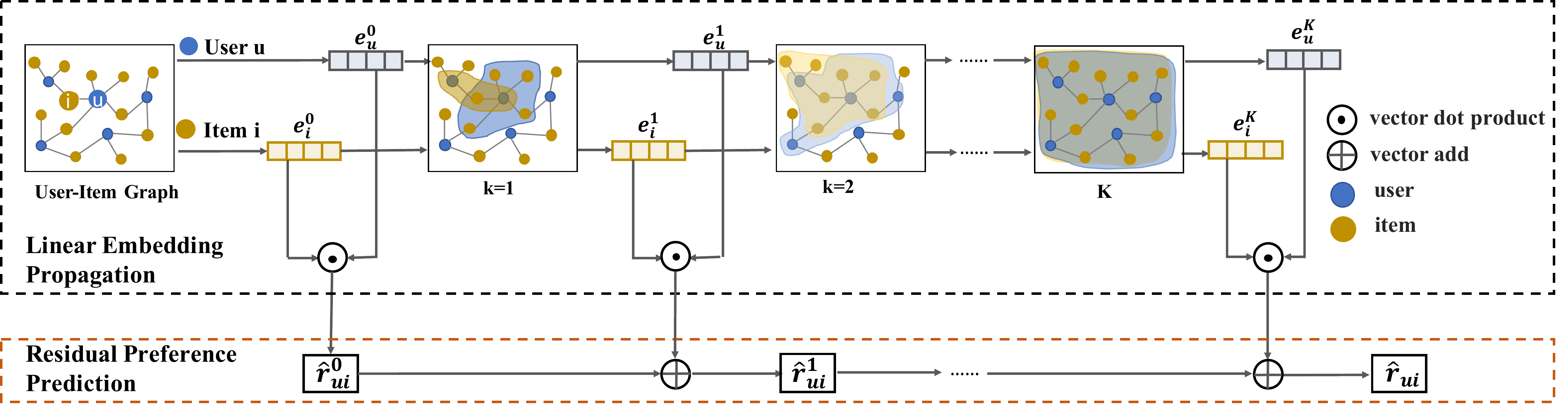}
	\end{center}
    \vspace{-0.5cm}
	\caption{The overall architecture of our proposed model.}\label{fig:framework}
   \vspace{-0.3cm}
\end{figure*} 
\end{small}

\subsection{Deep Network Architecture Design}
Theoretically, deep neural networks could approximate complex functions~\cite{goodfellow2016deep}. However, many researchers found stacking deeper layers in the network usually would not correspondingly increase performance in practice. For example, in the computer vision domain, directly stacking more layers in CNNs would complex the model training process, which leads to degradation of the image classification performance. 
For example, many CNNs variants have been proposed to how to stack more deep layers to improve image classification performance.~\cite{CVPR2016resudial,CVPR2017densely}. Researchers argued that the degradation of the deeper layers in CNNs is not caused by overfitting, but the harder training process with higher training error compared to the relatively shallower models.  Therefore, a deep residual learning framework, i.e., ResNet, is proposed to reformulate the layers as learning residual functions, which is easier to train compared to directly learning original functions~\cite{CVPR2016resudial}. In CF based recommender systems, simply relying on the deep neural networks would also not perform well due to the sparseness of user behavior data. Therefore, many deep learning based CF models have two parts: a shallow wide part and a deep neural network part, such as NeuMF~\cite{WWW2017neural} and Wide\&Deep~\cite{recsys2016wide}. The deep architecture design problem also exists in GCN variants. For example, many GCN based models achieve the best performance with layer depth of 2~\cite{NIPS2017inductive,wu2019neural}. As the local network structure varies from node to node, researchers proposed to aggregate all layer representations at the last layer~\cite{ICML2018representation}, or allowed the root node teleport to the later layers~\cite{klicpera2018predict}. In order to overcome the limitations of GCN models with limited labeled data, co-training and self-training approaches are proposed to train GCNs to supplement sparse labeled data~\cite{li2018deeper}. We differ from these works on two aspects. First, our model is based on a GCN with linear structure compared to these nonlinear GCNs. Moreover, our proposed architecture is concerned with how to better preserve the previous layer information with a residual network structure.

\section{Linear Residual Graph Convolutional Collaborative Filtering}

\subsection{Overall Structure of the Proposed Model}
In this part, we propose  \emph{L}inear \emph{R}esidual \emph{G}raph \emph{C}onvolutional \emph{C}ollaborative \emph{F}iltering~(LR-GCCF) which is a general GCN based CF model for recommendation. 
The overall architecture of LR-GCCF is shown in Figure~\ref{fig:framework}. LR-GCCF advances current GCN based models with two characteristics: (1)~At each layer of the feature propagation step, we use a simple linear embedding propagation without any nonlinear transformations. (2)~For predicting users' preferences of items, we propose a residual based network structure to overcome the limitations of previous works.

\subsubsection{Linear Embedding Propagation} 
Given the user-item bipartite graph as formulated in Eq.\eqref{eq:cfmat2graph}, let {\footnotesize $\mathbf{E}\in\mathbb{R}^{(M+N)\times D}$} denotes the free embeddings of users and items, with the first $M$ rows of the matrix, i.e.,~{\footnotesize $\Mat{E}_{[1:M]}$} is the user embedding submatrix, and~{\footnotesize $\Mat{E}_{[M+1:M+N]}$} is the item embedding submatrix. Then, LR-GCCF takes the embedding matrix as input:

\begin{small}
\vspace{-0.3cm}
\begin{equation}\label{eq:lrgcf_ini} 
\Mat{E}^0= \Mat{E},
\end{equation}
\vspace{-0.3cm}
\end{small}

\noindent which resembles the embedding based models in CF. Notably, different from GCN based tasks with node features as fixed input data, the embedding matrix is unknown and needs to be trained in LR-GCCF.

Following the theoretical elegance with graph spectral connections and empirical competing results of SGCN~\cite{ICML2019simplifying}, at each iteration step $k+1$, we assume the embedding matrix~{\footnotesize $\Mat{E}^{(k+1)}$}
is a linear aggregation of the embedding matrix~{\footnotesize $\Mat{E}^k$} at the previous layer $k$ as:

\begin{small}
\begin{equation} \label{eq:lrgcf_fp}
\Mat{E}^{k+1}=\mathbf{S}\Mat{E}^{k}\Mat{W}^k,
\end{equation}
\end{small}

\noindent where~{\small$\Mat{S}=\Mat{\tilde{D}}^{-0.5}\Mat{\tilde{A}}\Mat{\tilde{D}}^{-0.5}$} denotes the normalized adjacency matrix with added self loops,~{\footnotesize $\Mat{W}^k$} is the linear transformation. 

In fact, Eq.(\ref{eq:lrgcf_fp}) with matrix form is equivalent to modeling each user $u$'s and each item $i$'s updated embedding as:
\textbf{}
\begin{small}
\vspace{-0.2cm}
\begin{flalign}
[\mathbf{E}^{k+1}]_u=\mathbf{e}^{k+1}_u=[ \frac{1}{d_u}\mathbf{e}^k_u+\sum_{j\in R_u}\frac{1}{d_j\times d_u}\Mat{e}^k_j]\Mat{W}^k  \label{eq:lrgcf_fpu}\\
[\mathbf{E}^{k+1}]_i=\mathbf{e}^{k+1}_i=[\frac{1}{d_i}\mathbf{e}^k_i+\sum_{u\in R_i}\frac{1}{d_i\times d_u}\Mat{e}^k_u  ]\Mat{W}^k \label{eq:lrgcf_fpv},
\end{flalign}
% \vspace{-0.2cm}
\end{small}

\noindent which~$d_i$~($d_u$) is the diagonal degree of item~$i$~(user~$u$) in the user-item bipartite graph~$\mathcal{G}$. $R_*$ is neighbors of node~($*$) in graph~$\mathcal{G}$.

\subsubsection{Residual Preference Prediction}
With a predefined depth {\small$K$}, the recursive linear embedding  propagation would stop at the $K$-th layer with output of the embedding matrix~{\footnotesize $\Mat{E}^K$}. For each user~(item),~{\footnotesize $\Mat{e}^K_u$~($\Mat{e}^K_i$)} captures the up to K-th order bipartite graph similarity. Then, many embedding based recommendation models would predict the preference $\hat{r}_{ui}$ as the inner product between user and item latent vectors as:

\begin{small}
\vspace{-0.2cm}
\begin{equation}\label{eq:pred_r}
\hat{r}_{ui}=<\Mat{e}^K_u, \Mat{e}^K_i>,
\end{equation}
\vspace{-0.3cm}
\end{small}

\noindent where $<, >$ denotes vector inner product operation.

\begin{small}
\begin{figure} [htb]
	\begin{center}
		\includegraphics[width=.95\columnwidth]{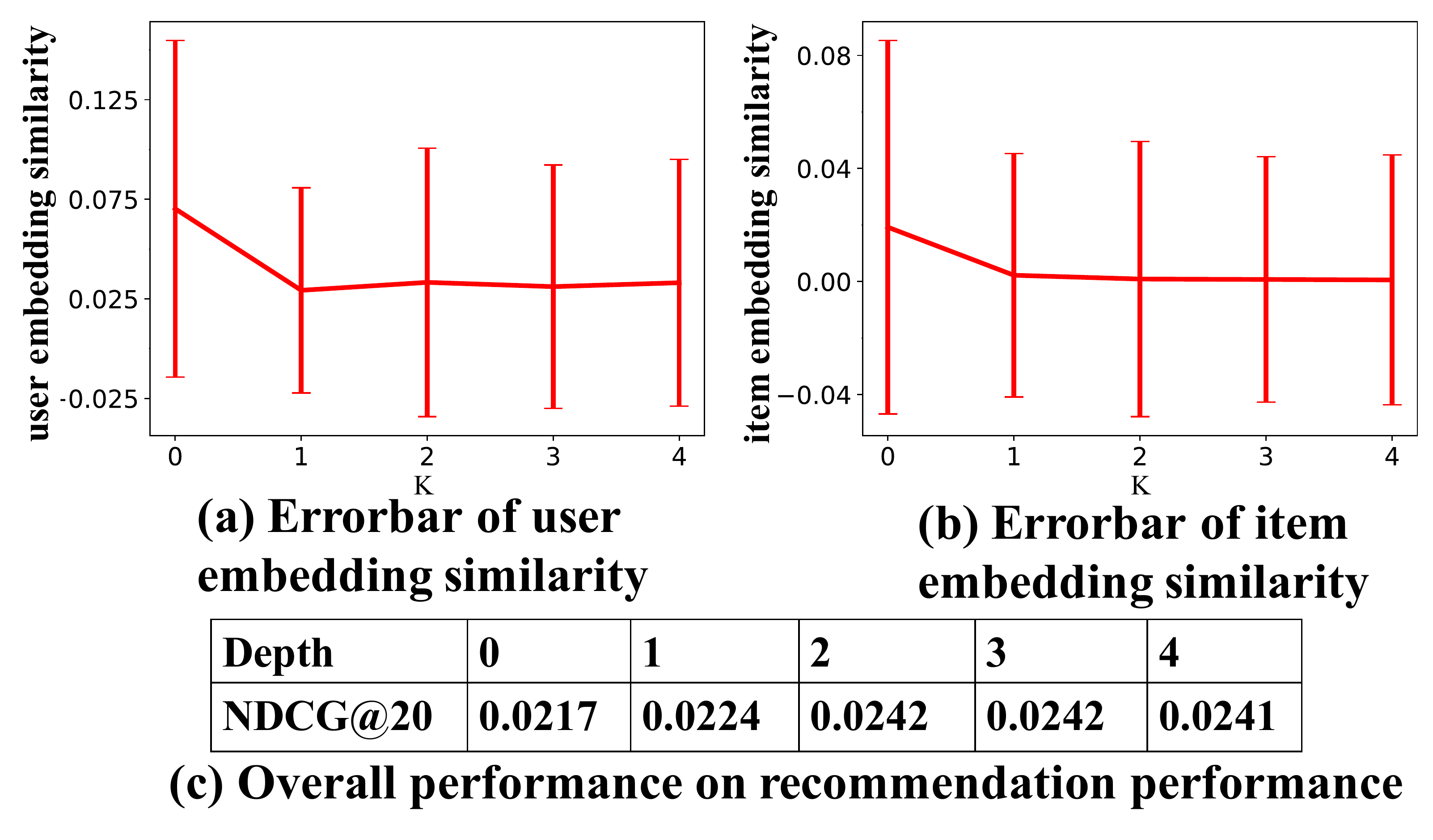}
	\end{center}
    \vspace{-0.5cm}
	\caption{GCN based recommendation performance with different layers $K$ on Amazon Books dataset.}\label{fig:k_performance}
   \vspace{-0.3cm}
\end{figure}
\end{small}

In practice, most GCN based variants, as well as GCN based recommendation models, achieve the best performance with {\small$K\!=\!2$}~\cite{ICLR2017semi,NIPS2017inductive,kdd2018Pinsage}. The overall trend for these GCN variants is that: the performance increases as {\small$K$} increases from 0 to 1~(2), and drops quickly as {\small$K$} continues to increase, the performance drops quickly. We speculate a possible reason is that, at the $k$-th layer, the embedding of each node is smoothed by the k-th order neighbors in the bipartite graph. Therefore, as $k$ increases from 0 to $K$, the node embeddings at deeper layers tend to be over smoothed, i.e.,  they are more similar with less distinctive information. This problem not only exists in GCNs, but is much more severe in CF with very sparse user behavior data for model learning. To validate the over assumption, we show the performance of GCN based recommendation with user-item bipartite graph using the predicted function in Eq.\eqref{eq:pred_r} with different depth {\small$K$}. When {\small$K\!=\!0$}, the GCN based recommendation model degenerates to BPR~\cite{UAI2009bpr}. To empirically show the over smoothing hypothesis, with each value of {\small$K$}, we calculate the average pair-wise user-user~(item-item) embedding similarity with cosine similarity at the $K$-th layer output. Specifically, for each pair of user $a$ and user $b$, their similarity is calculated as: $sim(a,b)\!=\! \frac{<\mathbf{e}^K_a, \mathbf{e}^K_b>}{||\mathbf{e}^K_a ||^2, ||\mathbf{e}^K_b||^2}$. Then, we plot the mean and variance of the cosine similarity of all pairs  in Figure~\ref{fig:k_performance}, with the recommendation performance is listed at the bottom. We have two observations from this figure. First, the variance between user~(item) embeddings are smaller when $K$ increases, due to the fact of the up to $K$-th order smoothness with neighborhood regularization. Second, when $K\!=\!0$, the recommendation performance is rather good. As we increase {\small$K$} from 0 to 2, the performance increases less than 10\%. Therefore, we empirically conclude that BPR~({\small$K\!=\!0$}) could already approximate preference of user to a large extent.

Based on the above two observations, we argue that: instead of directly approximating the user preference of each user-item pair at each layer, we perform the residual preference learning as:

\begin{small}
\vspace{-0.2cm}
\begin{equation}\label{eq:pred_rresidual}
\hat{r}^{k+1}_{ui}=\hat{r}^k_{ui}+<\Mat{e}^{k+1}_u,\Mat{e}^{k+1}_i>.
\end{equation}
\vspace{-0.2cm}
\end{small}

We hypothesis that it is easier to optimize the residual rating than to optimize the original rating, and the residual learning could help to alleviate the over smoothing effect with deeper layers.

Based on the residual preference prediction in above Eq.\eqref{eq:pred_rresidual}, we have:

\begin{small}
\vspace{-0.2cm}
\begin{flalign} \label{eq:pred_rvec}
\hat{r}_{ui}&=\hat{r}^{K-1}_{ui}+<\Mat{e}_u^{K}, \Mat{e}_i^{K}> \nonumber \\
&=\hat{r}^{K-2}_{ui}+ <\Mat{e}_u^{K-1}, \Mat{e}_i^{K-1}>+ <\Mat{e}_u^{K}, \Mat{e}_i^{K}> \nonumber \\
&=\hat{r}^{0}_{ui}+ <\Mat{e}_u^{1}, \Mat{e}_i^{1}> +... +<\Mat{e}_u^{K}, \Mat{e}_i^{K}> \nonumber \\
&=<\Mat{e}_u^{0}||\Mat{e}_u^{1}||...||\Mat{e}_u^{K}, \quad \Mat{e}_i^{0}||\Mat{e}_i^{1}||...||\Mat{e}_i^{K} >.
\end{flalign}
\vspace{-0.3cm}
\end{small}

The above equation is equivalent to concatenate embedding of each layer to form the final embedding of each node. This is quite reasonable as each node's sub-graph varies, and recording each layer's representation to form the final embedding of each node is more informative.

\subsubsection{Model Learning}
By putting the linear embedding propagation equation~(Eq.\eqref{eq:lrgcf_fp}) into vector representation of the 
residual prediction function~(Eq.\eqref{eq:pred_rvec}), we have:

\begin{small} 
\begin{flalign} \label{eq:lrgcf_wide}
\hat{r}_{ui}=&<\Mat{e}_u^{0}||\Mat{e}_u^{1}||...||\Mat{e}_u^{K},\quad \Mat{e}_v^{0}||\Mat{e}_v^{1}||...||\Mat{e}_v^{K} > \nonumber \\
=&<[\Mat{E}^{0}]_u||...||[\Mat{S}^K\Mat{E}^{0}\Mat{W}^0...\Mat{W}^K]_u,\nonumber·\\
&~~\quad  [\Mat{E}^{0}]_i||...||[\Mat{S}^K\Mat{E}^{0}\Mat{W}^0...\Mat{W}^K]_i> \nonumber \\
=&<[\Mat{E}^{0}]_u||...||[\Mat{S}^K\Mat{E}^{0}\Mat{Y}^K]_u, \quad  [\Mat{E}^{0}]_i||...||[\Mat{S}^K\Mat{E}^{0}\Mat{Y}^K]_i>,
\end{flalign} 
\end{small}

\noindent where $\Mat{Y}^K$ is reparameterized as $\Mat{Y}^K\!=\!\Mat{W}^0\Mat{W}^1...\Mat{W}^K$ with linear multiplication.~$\Mat{S}^K$ denotes the $K$-th power of~$\Mat{S}$.

Since we focus on implicit feedbacks,  we adopt the  pair-wise ranking  based loss function in BPR as:

\begin{small}
    \vspace{-0.2cm}
	\begin{equation}\label{eq:loss_r}
	\min\limits_{\Theta} \mathcal{L}(\mathbf{R},\mathbf{\hat{R}})=\sum_{a=1}^M\sum\limits_{(i,j)\in D_a } - ln(s(\hat{r}_{ai}-\hat{r}_{aj})) +\lambda||\Theta_1||^2,
	\end{equation}
	\vspace{-0.2cm}
\end{small}

\noindent where $s(x)$ is a sigmoid function. {\small $\Theta\!=\![\Theta_1,\Theta_2]$}, with {\small $\Theta_1\!=\![\Mat{E}^0]$}, and
{\small $\Theta_2\!=\![{[\mathbf{Y}^k]}_{k=1}^{K}]$}.  $\lambda$ is a regularization parameter that controls the complexity of user and item free embedding matrices. {\small$D_a=\{(i,j)|i\in R_a\!\wedge\!j\in V-R_a\}$} denotes the pairwise training data for $a$ with {\small$R_a$} represents the itemset that $a$ positively shows feedback.

\subsection{Model Discussion}

\textbf{Detailed Analysis of The Proposed Model.}~Based on the prediction function in Eq.\eqref{eq:lrgcf_wide}, we observe that LR-GCCF is not a deep neural network but a wide linear model. The linearization has several advantages: First, as LR-GCCF is built on the recent progress of SGC~\cite{ICML2019simplifying}, it is theoretically connected as a low pass filter of graph on the spectral domain. Second, with the linear embedding propagation and residual preference learning, LR-GCCF is much easier to train compared to nonlinear GCN based models. Last but not least, as our model does not have any hidden layers compared to deep learning based models, we do not need back propagation training algorithms. Instead, we could resort to the stochastic gradient descent for model learning. Therefore, LR-GCCF is much more time efficient compared to classical GCN based models.

\begin{footnotesize}
\begin{small}
\begin{table}[htb] \centering
\vspace{-0.5cm} 
\caption{Comparisons of different graph based recommendation models.}\label{tab:models_cmp}
\resizebox{0.95\columnwidth}{!}{
\begin{tabular}{|p{1.6cm}|p{0.7cm}|p{0.7cm}|p{1.5cm}|p{1.3cm}|} 
\hline
\multirow{3}{*}{Model} & \multicolumn{2}{c}{Graph Structure}  \vline &\multicolumn{2}{c}{Model Property} \vline\\
 \cline{2-5}
& First & Higher & Linear  & Residual  \\
& order & order & Propagation & Prediction  \\\hline
GC-MC & $\surd$ & $\times$ & $\times$ & $\times$ \\ \hline
Pinsage & $\surd$ & $\surd$ & $\times$ & $\times$ \\ \hline
NGCF& $\surd$ & $\surd$ & $\times$ & $\surd$  \\ \hline
\emph{\textbf{LR-GCCF}} & $\surd$ & $\surd$ & $\surd$ & $\surd$ \\ \hline
\end{tabular}
}
\vspace{-0.4cm}
\end{table}
\end{small}
\end{footnotesize}

\textbf{Connections with Previous Works.}~We compare the key characteristics of our proposed model with three closely related GCN based recommendation models: GC-MC~\cite{kdd2018GCMC}, PinSage~\cite{kdd2018Pinsage}, and NGCF~\cite{sigir2019ngcf}. In Table~\ref{tab:models_cmp}, NGCF is one of the first few attempts that also uses a residual prediction function  by taking each user~(item)'s embedding as a concatenation of all layers' embeddings. However, the authors simply use this ``trick" without any detailed explanation. We empirically show the reason why taking the output of the last layer embedding fails for CF, and shows using residual prediction is equivalent to concatenate all the layer's embeddings as the final embedding of each node in the user-item bipartite graph. For PinSage, it has lower time complexity compared to its deep learning based counterparts~(e.g., GC-MC and NGCF) as this model designed a sampling technique in feature aggregation process.

\section{Experiments}

\subsection{Experimental Setup}

\textbf{Datasets.}
We conduct experiments on two publicly available datasets: Amazon Books~\footnote{http://jmcauley.ucsd.edu/data/amazon/index.html} and Gowalla~\cite{liang2016modeling}. We summarize the statistics of two datasets in Table~\ref{tab:data_stats}. In data preprocessing step, we remove users~(items) that have less than 10 interaction records. After that, we randomly select 80\% of the records for training, 10\% for validation and the remaining 10\% for test. 

\begin{table}[htb] \centering
\vspace{-0.6cm}
\caption{\small{The statistics of the datasets.}}\label{tab:data_stats} 
\begin{scriptsize}
\begin{tabular}{|l|c|c|c|c|}
\hline
Dataset  & Users  & Items  &Ratings  & Rating Density \\ \hline 
Amazon Books &  52,643 & 91,599 & 2,984,108  & 0.062\%  \\  \hline 
Gowalla  &29,859 & 40,981 & 1,027,370& 0.084\%\\  \hline
\end{tabular}
\end{scriptsize}
\vspace{-0.3cm}
\end{table}

\textbf{Evaluation Metrics and Baselines.}
Since we focus on recommending items to users, we use two widely adopted ranking metrics for top-N recommendation evaluation: HR@N and NDCG@N~\cite{chen2017attentive}. For each user, we select all unrated items as the negative items and combine them with the positive items the user likes in the ranking process. We compare our proposed LR-GCCF model with various state-of-the-art baselines, including the classical model BPR~\cite{UAI2009bpr}, three graph convolutional based recommendation models: GC-MC~\cite{ICLR2017graph}, PinSage~\cite{kdd2018Pinsage}, and NGCF~\cite{sigir2019ngcf}. NGCF differs from PinSage as it adopts the residual learning process.
Besides, in order to better verify the effectiveness of the linear and the residual learning part, we design two variants of the GC-MC: ~\emph{L}inear-GC-MC~(L-GC-MC), and \emph{R}esudial-GC-MC~(R-GC-MC), with \emph{L} denotes replacing the original non-linear transformation with linear embedding propagation, and \emph{R} denotes the preference prediction. For the baseline of NGCF, as illustrated in  Table\ref{tab:models_cmp}, it adopts the residual preference learning, and when varying the non-linear embedding propagation to linear propagation, i.e., L-NGCF is the same as LR-GCCF, so we do not design variants of NGCF. For our proposed model LR-GCCF, we design a simplified version of ~\emph{L}inear-GCCF~(L-GCCF). In L-GCCF, we remove the residual learning process.

\textbf{Parameter Settings.} We implement our LR-GCCF model in Pytorch. There are two important parameters in our proposed model: the dimension D of the user and item embedding matrix~$\Mat{E}$, and the regularization parameter~$\lambda$~in the objective function~(Eq.\ref{eq:loss_r}). The embedding size is fixed to 64 for all models. In our proposed LR-GCCF model, we try the regularization parameter~$\lambda$~in the range~$[0.0001, 0.001, 0.01, 0.1]$,
and find~$\lambda=0.01$~reaches the best performance. We initialize the model parameters with a Gaussian distribution of mean 0 and standard deviation 0.01. There are several parameters in the baselines, for fair comparison, all the parameters in the baselines are also tuned to achieve the best performance.  For our proposed model, we empirically find that $\Mat{Y}$ equals the identity matrix, i.e., each parameter in $\Theta_2$ is not learned but directly set as the identity matrix reaches the best performance.

\subsection{Overall Comparison}

\begin{small} 
\begin{table*}[h]
\footnotesize
\centering
\caption{Performance of HR@N and NDCG@N on Amazon Books dataset. }\label{tab:resAmazon} 
\begin{tabular}{|l|l|l|l|l|l|l|l|l|l|l|}
\hline
\multicolumn{1}{|c|}{\multirow{2}{*}{Models}} & \multicolumn{2}{c|}{N=10} & \multicolumn{2}{c|}{N=20} & \multicolumn{2}{c|}{N=30} & \multicolumn{2}{c|}{N=40} & \multicolumn{2}{c|}{N=50} \\ \cline{2-11} 
\multicolumn{1}{|c|}{} & HR & NDCG & HR & NDCG & HR & NDCG & HR & NDCG & HR & NDCG \\ \hline
BPR & 0.01851 & 0.01710 & 0.02853 & 0.02169 & 0.03821 & 0.02564 & 0.04737 & 0.02911 & 0.05556 & 0.03205 \\ \hline 
GC-MC & 0.02063 & 0.01898 & 0.03196 & 0.02408 & 0.04242 & 0.02835 & 0.05226 & 0.03206 & 0.06133 & 0.03532 \\ \hline
PinSage & 0.02043 & 0.01872 & 0.03210 & 0.02404 & 0.04298 & 0.02844 & 0.05239 & 0.03199 & 0.06165 & 0.03529 \\ \hline
NGCF & 0.02071 & 0.01892 & 0.03244 & 0.02425 & 0.04343 & 0.02872 & 0.05329 & 0.03243 & 0.06263 & 0.03576 \\ \hline
\emph{L-GC-MC} & 0.02092 & 0.01916 & 0.03248 & 0.02443 & 0.04355 & 0.02894 & 0.05394 & 0.03286 & 0.06335 & 0.03623 \\ \hline
\emph{R-GC-MC} & 0.01962 & 0.01796 & 0.03084 & 0.02307 & 0.04153 & 0.02742 & 0.05139 & 0.03115 & 0.06032 & 0.03434 \\ \hline 
\emph{L-GCCF} &0.02067 &0.01909 &0.03200 &0.02424 &0.04312 &0.02876 &0.05310 &0.03254 &0.06218 &0.03579  \\ \hline
\emph{\textbf{LR-GCCF}} & \textbf{0.02209} & \textbf{0.02040} & \textbf{0.03407} & \textbf{0.02583} & \textbf{0.04532} & \textbf{0.03039} & \textbf{0.05532} & \textbf{0.03416} & \textbf{0.06498} & \textbf{0.03761} \\ \hline
\end{tabular} 
\end{table*} 
\end{small}

\begin{small} 
\begin{table*}[h]
\footnotesize
\centering
\caption{Performance of HR@N and NDCG@N on Gowalla dataset.}\label{tab:resGowalla} 
\begin{tabular}{|l|l|l|l|l|l|l|l|l|l|l|}
\hline
\multicolumn{1}{|c|}{\multirow{2}{*}{Models}} & \multicolumn{2}{c|}{N=10} & \multicolumn{2}{c|}{N=20} & \multicolumn{2}{c|}{N=30} & \multicolumn{2}{c|}{N=40} & \multicolumn{2}{c|}{N=50} \\ \cline{2-11} 
\multicolumn{1}{|c|}{} & HR & NDCG & HR & NDCG & HR & NDCG & HR & NDCG & HR & NDCG \\ \hline
BPR & 0.1041 & 0.1011 & 0.1378 & 0.1126 & 0.1664 & 0.1221 & 0.1908 & 0.1299 & 0.2122 & 0.1365 \\ \hline 
GC-MC & 0.1042 & 0.1010 & 0.1388 & 0.1127 & 0.1701 & 0.1222 & 0.1969 & 0.1307 & 0.2213 & 0.1381 \\ \hline
PinSage & 0.1057 & 0.1042 & 0.1390 & 0.1153 & 0.1682 & 0.1250 & 0.1935 & 0.1330 & 0.2146 & 0.1395 \\ \hline
NGCF & 0.1083 & 0.1094 & 0.1403 & 0.1197 & 0.1679 & 0.1288 & 0.1931 & 0.1368 & 0.2142 & 0.1432 \\ \hline
\emph{L-GC-MC} & 0.1045 & 0.1010 & 0.1399 & 0.1132 & 0.1701 & 0.1234 & 0.1957 & 0.1316 & 0.2184 & 0.1386 \\ \hline
\emph{R-GC-MC} & 0.1034 & 0.1000 & 0.1391 & 0.1123 & 0.1690 & 0.1224 & 0.1941 & 0.1305 & 0.2163 & 0.1373 \\ \hline 
\emph{L-GCCF} & 0.1044 & 0.1007 & 0.1412 & 0.1135 & 0.1721 & 0.1240 & 0.1977 & 0.1322 & 0.2196 & 0.1390 \\ \hline
\emph{\textbf{LR-GCCF}} & \textbf{0.1148} & \textbf{0.1136} & \textbf{0.1518} & \textbf{0.1259} & \textbf{0.1836} & \textbf{0.1365} & \textbf{0.2113} & \textbf{0.1453} & \textbf{0.2355} & 0\textbf{.1527} \\ \hline
\end{tabular} 
\end{table*} 
\end{small}

Table~\ref{tab:resAmazon}~and Table~\ref{tab:resGowalla} report the overall performance comparison results on HR@N and NDCG@N. GC-MC, PinSage, and NGCF improve over BPR by leveraging the user-item bipartite graph information. In particular, GC-MC and PinSage show the effectiveness of modeling the information passing of a graph.  NGCF is the baseline that captures higher-order user-item bipartite graph structure. It performs better than most baselines. Our proposed LR-GCCF model consistently outperforms NGCF, thus showing the effectiveness of modeling the user preference by the residual preference prediction and the linear embedding propagation.

In our proposed~LR-GCCF, the linear embedding propagation and residual preference learning are essential parts. To gain the effectiveness of these parts, we study the performance of the variants of baselines and our simplified model of~L-GC-MC.
We first analyze the performance of the linear embedding propagation by comparing the linear embedding based models with the counterparts that use non-linear embeddings, i.e., L-GC-MC vs. GC-MC. We find L-GC-MC outperforms GC-MC to a large margin, and similar trends exist
when comparing LR-GCCF and NGCF, empirically showing the effectiveness of the linear embedding propagation compared to the non-linear embedding propagation for GCN based recommendations. Next, we compare the performance of residual learning by comparing 
the results of  R-GC-MC vs GC-MC, the results of NGCF vs. PinSage, and the results of LR-GCCF and L-GCCF.  R-GC-MC does not show comparable performance as GC-MC, we guess a possible reason is that GC-MC is based on the first-order neighborhood aggregation. For the first-order neighborhood, each neighbor has limited neighbors and the over smoothing effect does not apply with first-order neighbors. With deep layers, the over smoothing effect  becomes more severe. Thus, NGCF outperforms PinSage, and LR-GCCF outperforms L-GCCF when modeling higher-order graph structure with residual learning. 
Last but not least, by combing the linear propagation and the residual learning together in LR-GCCF, the proposed model outperforms all the remaining models, showing the effectiveness of fusing these two parts for CF.

Instead of the nonlinear transformation of  feature propagation, our work differs from these works in a linearization method to accelerate the training process at the same time. In practice, we find that LR-GCCF is very easy to train. On Amazon Books dataset, with the best depth $K$ for each graph based recommendation model, at each iteration, the average runtime is about 30s for GC-MC~($K$=1), and 38s for PinSage~($K$=2) and NGCF~($K$=2), and about 20s for our proposed LR-GCCF~($K$=4) on a Ubuntu server with a single GTX 1080Ti. With larger K-th order graph embedding propagations, LR-GCCF costs less time with the linear embedding propagation. The runtime time on the Gowalla dataset for each model is about one third of the time compared to the time cost of the Amazon Books, and the overall trend of the time comparison is similar as analyzed above.

\begin{small}  
\begin{table}[] 
\caption{ Performance of HR@20 and NDCG@20 with different depth K.}\label{tab:diffK} 
\begin{tabular}{|l|l|l|l|l|}
\hline
\multicolumn{1}{|c|}{\multirow{2}{*}{Depth K}} & \multicolumn{2}{c|}{Amazon Books} & \multicolumn{2}{c|}{Gowalla} \\ \cline{2-5} 
\multicolumn{1}{|c|}{} & HR@20 & NDCG@20 & HR@20 & NDCG@20 \\ \hline
K=0 & 0.0285 &0.0217  &0.1378&0.1126 \\ \hline
K=1 & 0.0317 & 0.0242 & 0.1504 & 0.1246 \\ \hline
K=2 & 0.0327 & 0.0248 & 0.1506 & 0.1248 \\ \hline
K=3 & 0.0337 & 0.0255 & \textbf{0.1518} & \textbf{0.1259} \\ \hline
K=4 & \textbf{0.0341} & \textbf{0.0258} & 0.1494 & 0.1241 \\ \hline
K=5 & 0.0340 & 0.0257 & 0.1504 & 0.1247 \\ \hline
\end{tabular}
\end{table}  
\end{small}

\subsection{Detailed Model Analysis}
We would analyze the influence of the recursive label propagation depth $K$, and a detailed analysis of the learned embeddings of the residual preference prediction in~LR-GCCF.

Table~\ref{tab:diffK} shows the results on~LR-GCCF~with different K values. Particularly, the layer-wise propagation part disappears when $K$=0, i.e., our proposed model degenerates to BPR. As can be observed from Table~\ref{tab:diffK}, when K increase from 0 to 1, the performance increases quickly on both datasets. For Amazon Books, the best performance reaches with four propagation~depth. Meanwhile, our model reaches the best performance when $K$=3 on Gowalla.

\begin{small} 
\begin{figure} [htb]
	\begin{center}
		\includegraphics[width=.95\columnwidth]{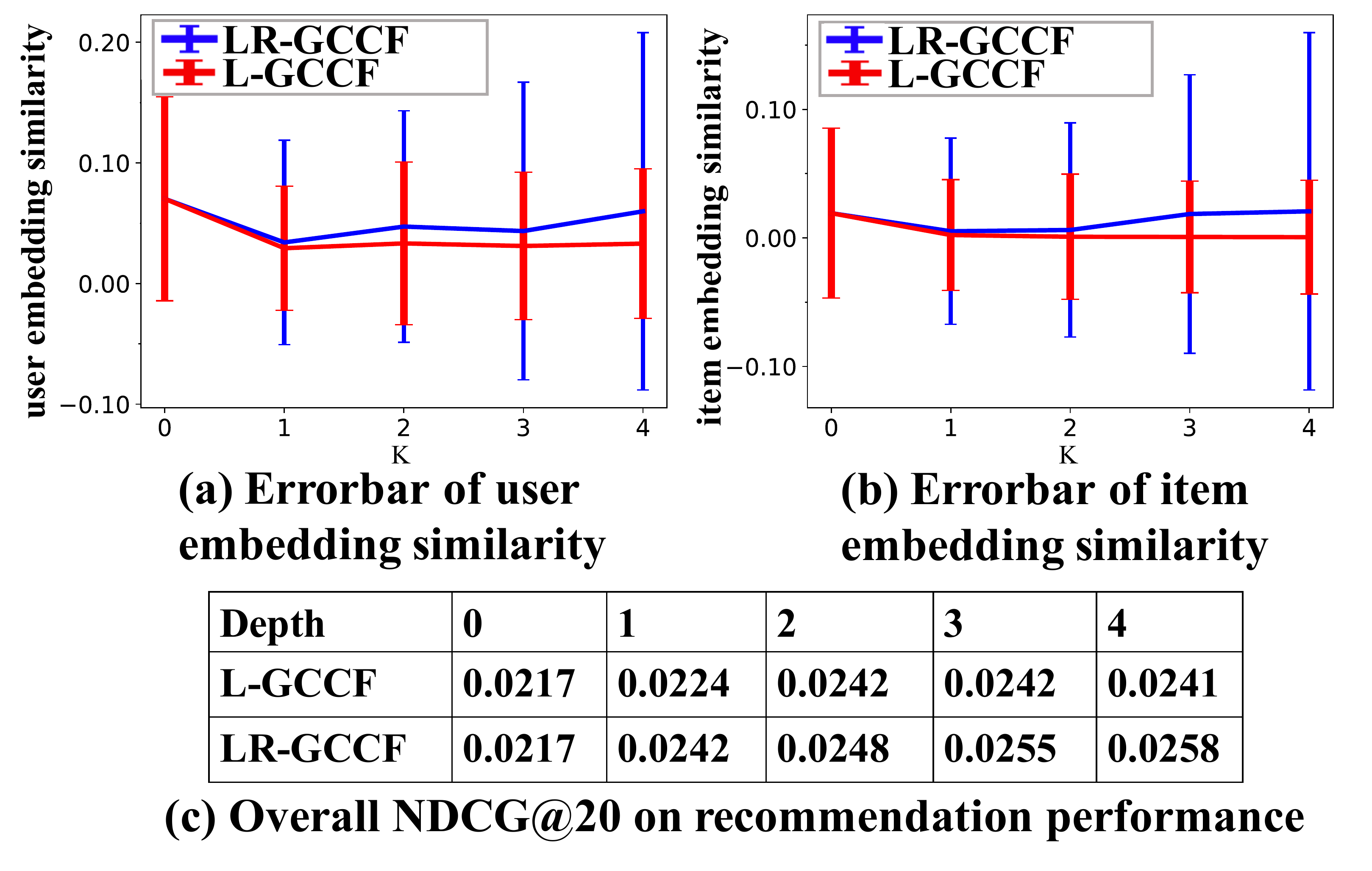}
	\end{center}
    \vspace{-0.5cm}
	\caption{Comparisons with and without residual preference prediction structure under different layers depth~$K$ on Amazon Books dataset.}\label{fig:sim_performance}
   \vspace{-0.3cm}
\end{figure}
\end{small}

In order to better show the effect of residual preference prediction, we design a simplified version of our proposed model that only removes the residual structure in our proposed model. We call the simplified model as L-GCCF. For L-GCCF and LR-GCCF, with each predefined depth $K$, we calculate the cosine similarity of each pair of users~(items) between their K-th layer output embedding, i.e., ~$\mathbf{e}^K$ for each node of the graph. The statistics of the mean and variance of user-user~(item-item) embedding similarities are shown in Figure~\ref{fig:sim_performance}. It obviously shows our proposed model has larger variance of the user-user cosine similarity compared to its counterparts L-GCCF that does not perform residual learning. This empirically validates that the residual learning could partially alleviate the over smoothing issue, and achieves better performance. Please note that, the overall trend on the Gowalla dataset is similar, and we do not show it due to page limit.

\section{Conclusions}
 In this paper, we revisited the current GCN based recommendation models, and proposed a LR-GCCF model for CF based recommendation. LR-GCCF was mainly composed of two parts: First,  with the recent progress of simple GCNs, we empirically removed the non-linear transformations in GCNs, and replaced it with linear embedding propagations. Second, to reduce the over smoothing effect introduced by higher layers of graph convolutions, we designed a residual preference prediction part with a residual preference learning process at each layer.  Extensive experimental results clearly showed the effectiveness and  efficiency of our proposed model. 
 In the future, we would like to explore how to better integrate the representations of different layers with well defined deep neural architectures  for better enhancing CF based recommendation.

\subsubsection{Acknowledgments.}
This work was supported in part by grants from  the National Key Research and Development Program of China~(2018YFB0804205), the National Natural Science Foundation of China~(Grant No. 61725203, 61972125, 61602147, 61932009, 61732008, 61722204), and Zhejiang Lab (No.2019KE0AB04).

\bibliographystyle{aaai}
\bibliography{aaaipaper} 

\begin{thebibliography}{}

\bibitem[\protect\citeauthoryear{Berg, Kipf, and Welling}{2018}]{kdd2018GCMC}
Berg, R. v.~d.; Kipf, T.~N.; and Welling, M.
\newblock 2018.
\newblock Graph convolutional matrix completion.
\newblock In {\em KDD Workshop}.

\bibitem[\protect\citeauthoryear{Camps-Valls, Marsheva, and
  Zhou}{2007}]{camps2007semi}
Camps-Valls, G.; Marsheva, T. V.~B.; and Zhou, D.
\newblock 2007.
\newblock Semi-supervised graph-based hyperspectral image classification.
\newblock {\em TGRS} 45(10):3044--3054.

\bibitem[\protect\citeauthoryear{Chen \bgroup et al\mbox.\egroup
  }{2017}]{chen2017attentive}
Chen, J.; Zhang, H.; He, X.; Nie, L.; Liu, W.; and Chua, T.-S.
\newblock 2017.
\newblock Attentive collaborative filtering: Multimedia recommendation with
  item-and component-level attention.
\newblock In {\em SIGIR},  335--344.

\bibitem[\protect\citeauthoryear{Cheng \bgroup et al\mbox.\egroup
  }{2016}]{recsys2016wide}
Cheng, H.-T.; Koc, L.; Harmsen, J.; Shaked, T.; Chandra, T.; Aradhye, H.;
  Anderson, G.; Corrado, G.; Chai, W.; Ispir, M.; et~al.
\newblock 2016.
\newblock Wide \& deep learning for recommender systems.
\newblock In {\em Recsys workshop},  7--10.

\bibitem[\protect\citeauthoryear{Gilmer \bgroup et al\mbox.\egroup
  }{2017}]{ICML2017neural}
Gilmer, J.; Schoenholz, S.~S.; Riley, P.~F.; Vinyals, O.; and Dahl, G.~E.
\newblock 2017.
\newblock Neural message passing for quantum chemistry.
\newblock In {\em ICML},  1263--1272.

\bibitem[\protect\citeauthoryear{Goodfellow, Bengio, and
  Courville}{2016}]{goodfellow2016deep}
Goodfellow, I.; Bengio, Y.; and Courville, A.
\newblock 2016.
\newblock {\em Deep learning}.
\newblock MIT press.

\bibitem[\protect\citeauthoryear{Gu, Zhou, and
  Ding}{2010}]{SDM2010collaborative}
Gu, Q.; Zhou, J.; and Ding, C.
\newblock 2010.
\newblock Collaborative filtering: Weighted nonnegative matrix factorization
  incorporating user and item graphs.
\newblock In {\em SDM},  199--210.

\bibitem[\protect\citeauthoryear{Hamilton, Ying, and
  Leskovec}{2017}]{NIPS2017inductive}
Hamilton, W.; Ying, Z.; and Leskovec, J.
\newblock 2017.
\newblock Inductive representation learning on large graphs.
\newblock In {\em NIPS},  1024--1034.

\bibitem[\protect\citeauthoryear{He \bgroup et al\mbox.\egroup
  }{2016}]{CVPR2016resudial}
He, K.; Zhang, X.; Ren, S.; and Sun, J.
\newblock 2016.
\newblock Deep residual learning for image recognition.
\newblock In {\em CVPR},  770--778.

\bibitem[\protect\citeauthoryear{He \bgroup et al\mbox.\egroup
  }{2017}]{WWW2017neural}
He, X.; Liao, L.; Zhang, H.; Nie, L.; Hu, X.; and Chua, T.-S.
\newblock 2017.
\newblock Neural collaborative filtering.
\newblock In {\em WWW},  173--182.

\bibitem[\protect\citeauthoryear{Huang \bgroup et al\mbox.\egroup
  }{2017}]{CVPR2017densely}
Huang, G.; Liu, Z.; Van Der~Maaten, L.; and Weinberger, K.~Q.
\newblock 2017.
\newblock Densely connected convolutional networks.
\newblock In {\em CVPR},  4700--4708.

\bibitem[\protect\citeauthoryear{Huang, Chung, and Chen}{2004}]{huang2004graph}
Huang, Z.; Chung, W.; and Chen, H.
\newblock 2004.
\newblock A graph model for e-commerce recommender systems.
\newblock {\em JASIST} 55(3):259--274.

\bibitem[\protect\citeauthoryear{Kipf and Welling}{2017}]{ICLR2017semi}
Kipf, T.~N., and Welling, M.
\newblock 2017.
\newblock Semi-supervised classification with graph convolutional networks.
\newblock In {\em ICLR}.

\bibitem[\protect\citeauthoryear{Klicpera, Bojchevski, and
  G{\"u}nnemann}{2019}]{klicpera2018predict}
Klicpera, J.; Bojchevski, A.; and G{\"u}nnemann, S.
\newblock 2019.
\newblock Predict then propagate: Graph neural networks meet personalized
  pagerank.
\newblock {\em ICLR}.

\bibitem[\protect\citeauthoryear{Koren, Bell, and
  Volinsky}{2009}]{koren2009matrix}
Koren, Y.; Bell, R.; and Volinsky, C.
\newblock 2009.
\newblock Matrix factorization techniques for recommender systems.
\newblock {\em Computer} 42(8):30--37.

\bibitem[\protect\citeauthoryear{Li \bgroup et al\mbox.\egroup
  }{2016}]{li2016unsupervised}
Li, D.; Hung, W.-C.; Huang, J.-B.; Wang, S.; Ahuja, N.; and Yang, M.-H.
\newblock 2016.
\newblock Unsupervised visual representation learning by graph-based consistent
  constraints.
\newblock In {\em ECCV},  678--694.

\bibitem[\protect\citeauthoryear{Li, Han, and Wu}{2018}]{li2018deeper}
Li, Q.; Han, Z.; and Wu, X.-M.
\newblock 2018.
\newblock Deeper insights into graph convolutional networks for semi-supervised
  learning.
\newblock In {\em AAAI}.

\bibitem[\protect\citeauthoryear{Liang \bgroup et al\mbox.\egroup
  }{2016}]{liang2016modeling}
Liang, D.; Charlin, L.; McInerney, J.; and Blei, D.~M.
\newblock 2016.
\newblock Modeling user exposure in recommendation.
\newblock In {\em WWW},  951--961.

\bibitem[\protect\citeauthoryear{Liu and Yang}{2008}]{SIGIR2008eigenrank}
Liu, N.~N., and Yang, Q.
\newblock 2008.
\newblock Eigenrank: a ranking-oriented approach to collaborative filtering.
\newblock In {\em SIGIR},  83--90.

\bibitem[\protect\citeauthoryear{Monti, Bronstein, and
  Bresson}{2017}]{NIPSi2017geometric}
Monti, F.; Bronstein, M.; and Bresson, X.
\newblock 2017.
\newblock Geometric matrix completion with recurrent multi-graph neural
  networks.
\newblock In {\em NIPS},  3697--3707.

\bibitem[\protect\citeauthoryear{Rendle \bgroup et al\mbox.\egroup
  }{2009}]{UAI2009bpr}
Rendle, S.; Freudenthaler, C.; Gantner, Z.; and Schmidt-Thieme, L.
\newblock 2009.
\newblock Bpr: Bayesian personalized ranking from implicit feedback.
\newblock In {\em UAI},  452--461.

\bibitem[\protect\citeauthoryear{van~den Berg, Kipf, and
  Welling}{2017}]{ICLR2017graph}
van~den Berg, R.; Kipf, T.~N.; and Welling, M.
\newblock 2017.
\newblock Graph convolutional matrix completion.
\newblock {\em KDD}.

\bibitem[\protect\citeauthoryear{Velickovic \bgroup et al\mbox.\egroup
  }{2018}]{ICLR2018graphATT}
Velickovic, P.; Cucurull, G.; Casanova, A.; Romero, A.; Lio, P.; and Bengio, Y.
\newblock 2018.
\newblock Graph attention networks.
\newblock In {\em ICLR}.

\bibitem[\protect\citeauthoryear{Wang \bgroup et al\mbox.\egroup
  }{2019}]{sigir2019ngcf}
Wang, X.; He, X.; Wang, M.; Feng, F.; and Chua, T.-S.
\newblock 2019.
\newblock Neural graph collaborative filtering.
\newblock In {\em SIGIR}.

\bibitem[\protect\citeauthoryear{Wu \bgroup et al\mbox.\egroup
  }{2016}]{wu2016relevance}
Wu, L.; Liu, Q.; Chen, E.; Yuan, N.~J.; Guo, G.; and Xie, X.
\newblock 2016.
\newblock Relevance meets coverage: A unified framework to generate diversified
  recommendations.
\newblock {\em TIST} 7(3):39.

\bibitem[\protect\citeauthoryear{Wu \bgroup et al\mbox.\egroup
  }{2017}]{wu2017modeling}
Wu, L.; Ge, Y.; Liu, Q.; Chen, E.; Hong, R.; Du, J.; and Wang, M.
\newblock 2017.
\newblock Modeling the evolution of users’ preferences and social links in
  social networking services.
\newblock {\em TKDE} 29(6):1240--1253.

\bibitem[\protect\citeauthoryear{Wu \bgroup et al\mbox.\egroup
  }{2019a}]{ICML2019simplifying}
Wu, F.; Zhang, T.; Souza~Jr, A. H.~d.; Fifty, C.; Yu, T.; and Weinberger, K.~Q.
\newblock 2019a.
\newblock Simplifying graph convolutional networks.
\newblock In {\em ICML},  6861--6871.

\bibitem[\protect\citeauthoryear{Wu \bgroup et al\mbox.\egroup
  }{2019b}]{wu2019neural}
Wu, L.; Sun, P.; Fu, Y.; Hong, R.; Wang, X.; and Wang, M.
\newblock 2019b.
\newblock A neural influence diffusion model for social recommendation.
\newblock In {\em SIGIR},  235--244.

\bibitem[\protect\citeauthoryear{Wu, Liu, and Yang}{2018}]{kdir18}
Wu, Y.; Liu, H.; and Yang, Y.
\newblock 2018.
\newblock Graph convolutional matrix completion for bipartite edge prediction.
\newblock In {\em KDIR},  51--60.

\bibitem[\protect\citeauthoryear{Wu, Shen, and Van
  Den~Hengel}{2019}]{wu2019wider}
Wu, Z.; Shen, C.; and Van Den~Hengel, A.
\newblock 2019.
\newblock Wider or deeper: Revisiting the resnet model for visual recognition.
\newblock {\em Pattern Recognition} 90:119--133.

\bibitem[\protect\citeauthoryear{Xu \bgroup et al\mbox.\egroup
  }{2018}]{ICML2018representation}
Xu, K.; Li, C.; Tian, Y.; Sonobe, T.; Kawarabayashi, K.-i.; and Jegelka, S.
\newblock 2018.
\newblock Representation learning on graphs with jumping knowledge networks.
\newblock In {\em ICML}.

\bibitem[\protect\citeauthoryear{Xu \bgroup et al\mbox.\egroup
  }{2019a}]{xu2019relation}
Xu, F.; Lian, J.; Han, Z.; Li, Y.; Xu, Y.; and Xie, X.
\newblock 2019a.
\newblock Relation-aware graph convolutional networks for agent-initiated
  social e-commerce recommendation.
\newblock In {\em CIKM},  529--538.

\bibitem[\protect\citeauthoryear{Xu \bgroup et al\mbox.\egroup
  }{2019b}]{ICLR2019powerful}
Xu, K.; Hu, W.; Leskovec, J.; and Jegelka, S.
\newblock 2019b.
\newblock How powerful are graph neural networks?
\newblock In {\em ICLR}.

\bibitem[\protect\citeauthoryear{Ying \bgroup et al\mbox.\egroup
  }{2018}]{kdd2018Pinsage}
Ying, R.; He, R.; Chen, K.; Eksombatchai, P.; Hamilton, W.~L.; and Leskovec, J.
\newblock 2018.
\newblock Graph convolutional neural networks for web-scale recommender
  systems.
\newblock In {\em SIGKDD},  974--983.

\bibitem[\protect\citeauthoryear{Zhao \bgroup et al\mbox.\egroup
  }{2019}]{zhao2019t}
Zhao, L.; Song, Y.; Zhang, C.; Liu, Y.; Wang, P.; Lin, T.; Deng, M.; and Li, H.
\newblock 2019.
\newblock T-gcn: A temporal graph convolutional network for traffic prediction.
\newblock {\em TITS}.

\bibitem[\protect\citeauthoryear{Zheng \bgroup et al\mbox.\egroup
  }{2018}]{zheng2018spectral}
Zheng, L.; Lu, C.-T.; Jiang, F.; Zhang, J.; and Yu, P.~S.
\newblock 2018.
\newblock Spectral collaborative filtering.
\newblock In {\em RecSys},  311--319.

\bibitem[\protect\citeauthoryear{Zhu, Ghahramani, and
  Lafferty}{2003}]{zhu2003semi}
Zhu, X.; Ghahramani, Z.; and Lafferty, J.~D.
\newblock 2003.
\newblock Semi-supervised learning using gaussian fields and harmonic
  functions.
\newblock In {\em ICML},  912--919.

\end{thebibliography}

\end{document}